\documentclass[epj]{webofc} 
\usepackage[varg]{txfonts}   

\woctitle{HYP2022 - 14$^\textrm{th}$ International Conference on Hypernuclear 
and Strange Particle Physics} 

\def\lamb#1#2{$^{#1}_{\Lambda}${#2}} 
\def\lam#1#2{$^{#1}_{~\Lambda}${#2}} 
\def\la#1#2{$^{#1}_{~~\Lambda}${#2}}

\usepackage{slashed} 

\begin{document} 

\title{$\Lambda NN$ content of $\Lambda$-nucleus potential} 
\author{\firstname{Eliahu} \lastname{Friedman}\inst{1}\fnsep
\thanks{Eliahu.Friedman@mail.huji.ac.il} \and \firstname{Avraham}
\lastname{Gal}\inst{1}\fnsep\thanks{avragal@savion.huji.ac.il}
\institute{Racah Institute of Physics, The Hebrew University, 
Jerusalem 9190401, Israel}} 

\abstract{A minimally constructed $\Lambda$-nucleus density-dependent 
optical potential is used to calculate binding energies of observed 
$1s_{\Lambda}$, $1p_{\Lambda}$ states across the periodic table, leading 
to a repulsive $\Lambda NN$ contribution $D_{\Lambda}^{(3)}\approx$14~MeV 
to the phenomenological $\Lambda$-nucleus potential depth $D_{\Lambda}
\approx -$30~MeV. This value is significant in connection with the so-called 
'hyperon puzzle.'}

\maketitle

\section{Introduction} 
\label{sec:intro} 

\begin{figure}[h] 
\centering 
\includegraphics[width=0.7\textwidth]{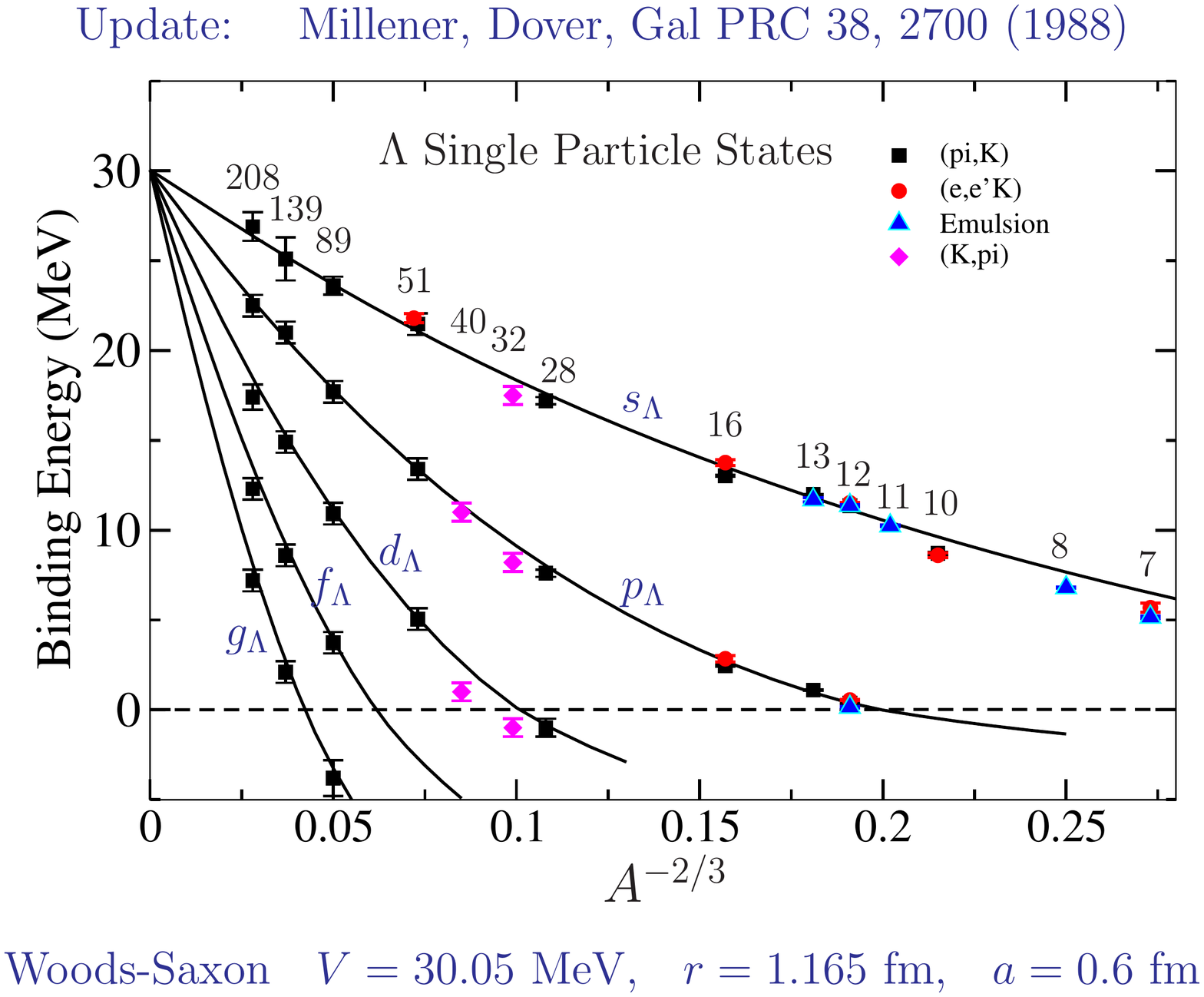} 
\caption{Compilation of $\Lambda$ binding energies in \lamb{7}{Li} to 
\la{208}{Pb} from various sources, and as calculated using a three-parameter 
WS potential \cite{MDG88}. Figure adapted from Ref. \cite{GHM16}} 
\label{fig:mdg} 
\end{figure}

The $\Lambda$-nucleus potential depth provides an important constraint in 
ongoing attempts to resolve the `hyperon puzzle', i.e., whether or not dense 
neutron-star matter contains hyperons, primarily $\Lambda$s besides nucleons 
\cite{FT20}. Figure~\ref{fig:mdg} presents compilation of most of the known 
$\Lambda$ hypernuclear binding energies ($B_\Lambda$) across the periodic 
table, fitted by a three-parameter Woods-Saxon (WS) attractive potential. 
As $A\to\infty$, a limiting value of $B_{\Lambda}(A)\to 30$~MeV is obtained. 
Interestingly, studies of density dependent $\Lambda$-nuclear optical 
potentials $V_{\Lambda}(\rho)$ in Ref.~\cite{MDG88}, with $\rho $ the nuclear 
density normalized to the number of nucleons A, conclude that a $\rho^2$ term 
motivated by three-body $\Lambda NN$ interactions provides a large repulsive 
(positive) contribution to the $\Lambda$-nuclear potential depth $D_{\Lambda}$ 
at nuclear-matter density $\rho_0$: $D_{\Lambda}^{(3)}\approx 30$~MeV. 
This repulsive component of $D_{\Lambda}$ is more than just compensated at 
$\rho_0$ by a roughly twice larger attractive depth value $D_{\Lambda}^{(2)}
\approx -60$~MeV, motivated by a two-body $\Lambda N$ interaction. Note that 
$D_{\Lambda}$ is defined as $V_{\Lambda}(\rho_0)$ in the limit $A\to\infty$ 
at a given nuclear-matter density $\rho_0$, with a value 0.17~fm$^{-3}$ 
assumed here. 

Most hyperon-nucleon potential models overbind $\Lambda$ hypernuclei, yielding 
values of $D_{\Lambda}^{(2)}$ deeper than $-30$~MeV. Whereas such overbinding 
amounts to only few MeV in the often used Nijmegen soft-core model versions 
NSC97e,f \cite{NSC97} it is considerably stronger, by more than 10~MeV, 
in the recent Nijmegen extended soft-core model ESC16 \cite{ESC16}. A similar 
overbinding arises at leading order in chiral effective field theory 
($\chi$EFT) \cite{LO06}. The situation at next-to leading order (NLO) is less 
clear owing to a strong dependence of $D_{\Lambda}^{(2)}$ on the momentum 
cutoff scale $\lambda$ \cite{HV20}. At $\lambda$=500~MeV/c, however, it is 
found in Ref.~\cite{GKW20} that both versions NLO13 \cite{NLO13} and NLO19 
\cite{NLO19} overbind by a few MeV. Finally, recent Quantum Monte Carlo (QMC) 
calculations \cite{LGP13,LPG14}, using a $\Lambda N+ \Lambda NN$ interaction 
model designed to bind correctly \lamb{5}{He}, result in a strongly attractive 
$D_{\Lambda}^{(2)}$ of order $-100$~MeV and a correspondingly large repulsive 
(positive) $D_\Lambda^{(3)}$, reproducing the overall potential depth 
$D_{\Lambda}\approx -30$ MeV. 

Our aim in the present phenomenological study is to check to what extent 
properly chosen $\Lambda$ hypernuclear binding energy data, with {\it minimal} 
extra assumptions, imply positive values of $D_\Lambda^{(3)}$, and how large 
it is~\cite{FG22}. Repulsive three-body $\Lambda NN$ interactions go beyond 
just providing solution of the overbinding problem: as nuclear density 
is increased beyond nuclear matter density $\rho_0$, the balance between 
attractive $D_\Lambda^{(2)}$ and repulsive $D_\Lambda^{(3)}$ tilts towards 
the latter. This results in nearly total expulsion of $\Lambda$ hyperons 
from neutron-star matter, suggesting an equation of state (EoS) sufficiently 
stiff to support two solar-mass neutron stars, thereby providing a possible 
solution to the `hyperon puzzle'. The larger $D_\Lambda^{(3)}$ is, the more 
likely it is a solution \cite{LLGP15,LVB19}. However, there is no guarantee 
that three-body $\Lambda NN$ interactions are universally repulsive. 
For a recent discussion of this problem within an SU(3) `decuplet dominance' 
approach practised in modern $\chi$EFT studies at NLO, see Ref.~\cite{GKW20}. 

In this Contribution we adopt the optical potential approach as applied 
by Dover-H\"{u}fner-Lemmer to pions in nuclear matter \cite{DHL71}. 
For the $\Lambda$-nucleus system, it provides expansion in powers 
of the nuclear density $\rho(r)$, consisting of a linear term induced by 
a two-body $\Lambda N$ interaction plus two higher-power density terms: 
(i) a long-range Pauli correlations term starting at $\rho^{4/3}$, and 
(ii) a short-range $\Lambda NN$ interaction term dominated in the present 
context by three-body $\Lambda NN$ interactions, starting at $\rho^2$. 
As demonstrated below, the contribution of the Pauli correlations term is 
non negligible, propagating to higher powers of density terms than just 
$\rho^{4/3}$, such as the $\rho^2$ $\Lambda NN$ interaction term. This 
explains why the value derived here, $D_\Lambda^{(3)}=(13.9\pm 1.4)$~MeV, 
differs from any of those suggested earlier in Ref.~\cite{MDG88} and in 
Skyrme Hartree Fock studies \cite{SH14} where Pauli correlations are usually 
disregarded. Our value of $D_\Lambda^{(3)}$ strongly disagrees with the much 
larger value inferred in QMC calculations \cite{LPG14}. We comment on these 
discrepancies below.

\section{Nuclear densities}
\label{sec:dens}

In optical model applications aimed at establishing relations between 
components with different powers of density $\rho=\rho_p+\rho_n$, it is 
crucial to ensure that the radial extent of the densities, e.g., their r.m.s. 
radii, follows closely values derived from experiment. For proton densities 
we used charge densities, with proton finite-size and recoil effects included. 
Harmonic-oscillator type densities~\cite{Elton61} were used for the lightest 
elements, assuming the same radial parameters for protons and neutrons. 
A variation of 1\% in the r.m.s. neutron radius was found to affect 
calculated $\Lambda$ binding energies considerably less than given by 
most of the experimental uncertainties listed in Table~\ref{tab:data} below. 
For a detailed discussion in the analogous case of light $\Xi^-$ hypernuclei, 
see Ref.~\cite{FG21}. For species beyond the nuclear $1p$ shell we used 
two-parameter Fermi distributions normalized to $Z$ for protons and $N=A-Z$ 
for neutrons, derived from assembled nuclear charge distributions~\cite{AM13}. 
For medium-weight and heavy nuclei, the r.m.s. radii of our neutron 
density distributions assume larger values than those for proton density 
distributions, as practiced in analyses of exotic atoms \cite{FG07}. 
Furthermore, once neutron orbits extend beyond proton orbits, 
it is useful to represent the nuclear density $\rho(r)$ as 
\begin{equation} 
\rho(r)=\rho_{\rm core}(r)+\rho_{\rm excess}(r), 
\label{eq:exc1} 
\end{equation} 
where $\rho_{\rm core}$ refers to the $Z$ protons plus the charge symmetric 
$Z$ neutrons occupying the same nuclear `core' orbits, and $\rho_{\rm excess}$ 
refers to the $(N-Z)$ `excess' neutrons associated with the nuclear periphery.

\section{Optical potential}
\label{sec:potls}

The optical potential employed in this work, 
$V_{\Lambda}^{\rm opt}(\rho)=V_{\Lambda}^{(2)}(\rho)+V_{\Lambda}^{(3)}(\rho)$, 
consists of terms representing two-body $\Lambda N$ and three-body 
$\Lambda NN$ interactions, respectively: 
\begin{equation} 
V_{\Lambda}^{(2)}(\rho) = -\frac{4\pi}{2\mu_{\Lambda}}f_A\,
C_{\rm Pauli}(\rho)\,b_0\rho, 
\label{eq:V2} 
\end{equation} 
\begin{equation} 
V_{\Lambda}^{(3)}(\rho) = +\frac{4\pi}{2\mu_{\Lambda}}\,f_A\,B_0\,
\frac{\rho^2}{\rho_0}, 
\label{eq:V3} 
\end{equation} 
with $b_0$ and $B_0$ strength parameters in units of fm ($\hbar=c=1$). 
In these expressions, $\rho(r)$ is a nuclear density distribution normalized 
to the number of nucleons $A$, $\rho_0=0.17$~fm$^{-3}$ stands for 
nuclear-matter density, $\mu_{\Lambda}$ is the $\Lambda$-nucleus reduced mass 
and $f_A$ is a kinematical factor transforming $b_0$ from the $\Lambda N$ c.m. 
system to the $\Lambda$-nucleus c.m. system: 
\begin{equation} 
f_A=1+\frac{A-1}{A}\frac{\mu_{\Lambda}}{m_N}. 
\label{eq:fA} 
\end{equation} 
This form of $f_A$ coincides with the way it is used for $V_{\Lambda}^{(2)}$ 
in atomic/nuclear hadron-nucleus bound-state problems~\cite{FG07} and its $A$ 
dependence provides good approximation for $V_{\Lambda}^{(3)}$. Next is the 
density dependent factor $C_{\rm Pauli}(\rho)$ in Eq.~(\ref{eq:V2}), standing 
for a Pauli correlation function: 
\begin{equation} 
C_{\rm Pauli}(\rho)=(1+\alpha_P\frac{3k_F}{2\pi}f_Ab_0)^{-1}, 
\label{eq:Cpauli1} 
\end{equation} 
with Fermi momentum $k_F=(3{\pi^2}\rho/2)^{1/3}$. The parameter $\alpha_P$ 
in Eq.~(\ref{eq:Cpauli1}) switches off ($\alpha_P$=0) or on ($\alpha_P$=1) 
Pauli correlations in a form suggested in Ref.~\cite{WRW97} and practised 
in $K^-$ atoms studies \cite{FG17}. To estimate $1/A$ correction terms, 
we also approximated $C_{\rm Pauli}(\rho)$ by~\cite{FG21}:  
\begin{equation} 
C_{\rm Pauli}(\rho)\approx(1+\alpha_P\frac{3k_F}{2\pi}b_0^{\rm lab})^{-1}, 
\,\,\,\,\,\,\,\,\,\, b_0^{\rm lab}=(1+\frac{m_{\Lambda}}{m_N})\,b_0.  
\label{eq:Cpauli2} 
\end{equation} 
As shown below, including $C_{\rm Pauli}(\rho)$ in $V_{\Lambda}^{(2)}$ affects 
strongly the balance between the derived potential depths $D_\Lambda^{(2)}$ 
and $D_\Lambda^{(3)}$. However, introducing it also in $V_{\Lambda}^{(3)}$ 
is found to make little difference, which is why it is skipped in 
Eq.~(\ref{eq:V3}). Finally we note that the low-density limit of 
$V_{\Lambda}^{\rm opt}$ requires according to Ref.~\cite{DHL71} that $b_0$ 
is identified with the c.m. $\Lambda N$ spin-averaged scattering length 
(positive here).

\section{Data} 
\label{sec:data} 

\begin{table}[htb]
\begin{center}
\caption{Binding energies in MeV, including uncertainties, considered here; 
taken from Table IV of Ref.~\cite{GHM16}.} 
\label{tab:data}       
\centering 
\begin{tabular}{cllll} 
\hline 
hypernucleus & ~~$1s_\Lambda $ & $~~\pm$ & $1p_\Lambda $ & $~~\pm $ \\ \hline 
$^{12}_{~\Lambda}$B & 11.52 & 0.02 & 0.54  & 0.04  \\ 
$^{13}_{~\Lambda}$C & 12.0  & 0.2  & 1.1   & 0.2   \\ 
$^{16}_{~\Lambda}$N & 13.76  & 0.16  & 2.84   & 0.18   \\ 
$^{28}_{~\Lambda}$Si & 17.2  & 0.2  & 7.6   & 0.2   \\ 
$^{32}_{~\Lambda}$S & 17.5  & 0.5  & 8.2   & 0.5   \\ 
$^{51}_{~\Lambda}$V & 21.5  & 0.6  & 13.4   & 0.6   \\ 
$^{89}_{~\Lambda}$Y & 23.6  & 0.5  & 17.7   & 0.6   \\ 
$^{139}_{~~\Lambda}$La & 25.1  & 1.2  & 21.0   & 0.6   \\ 
$^{208}_{~~\Lambda}$Pb & 26.9  & 0.8  & 22.5   & 0.6   \\  \hline 
\end{tabular} 
\end{center} 
\end{table} 

\begin{figure}[h] 
\centering 
\includegraphics[width=0.7\textwidth]{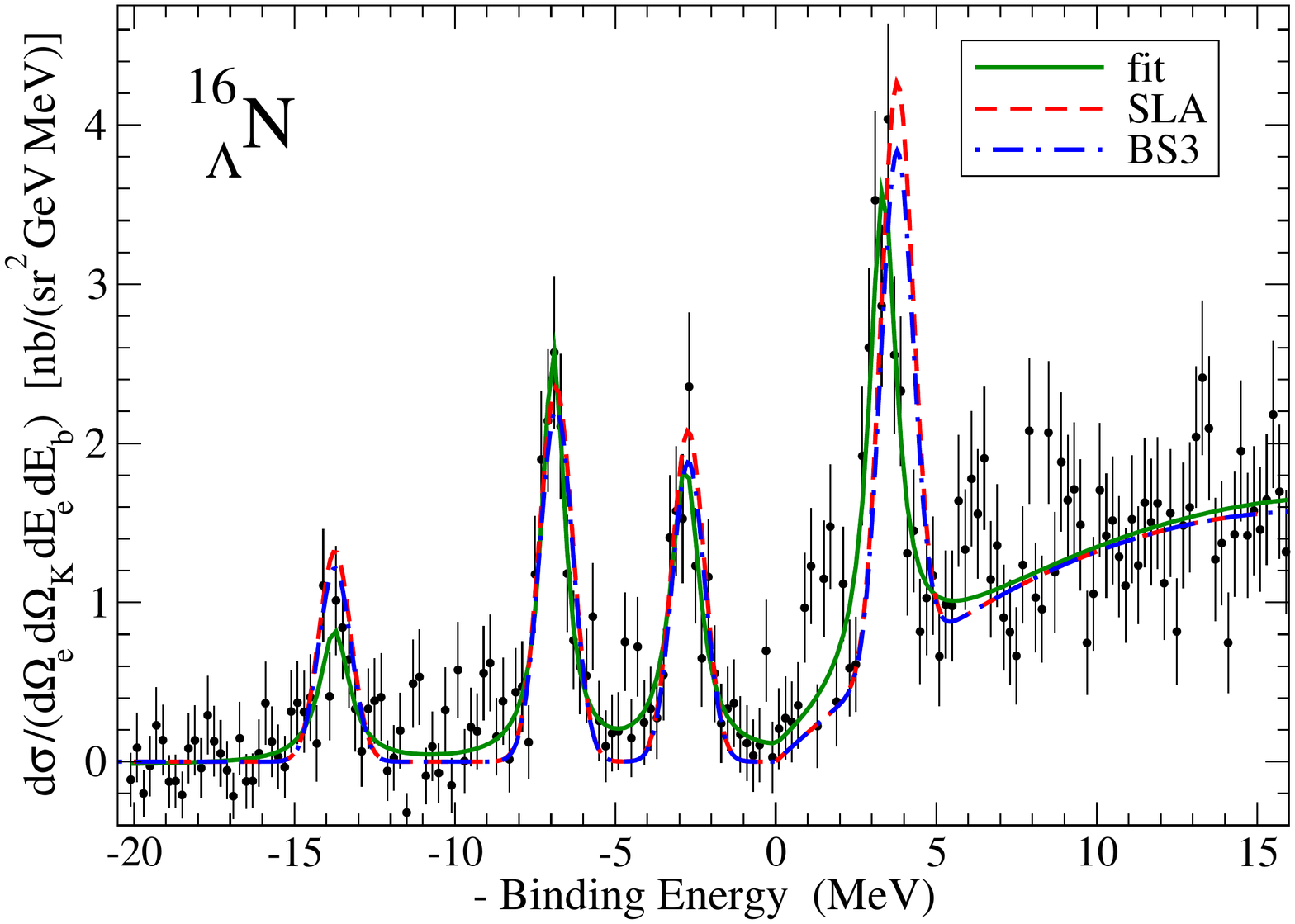} 
\caption{$^{16}$O($e,e'K^+$) spectrum of \lam{16}{N} from JLab Hall A 
measurements. Figure adapted from Ref.~\cite{JLAB19}.} 
\label{fig:L16N} 
\end{figure}

The present work does not attempt to reproduce the full range of $B_{\Lambda}$ 
data shown in Fig.~\ref{fig:mdg}. It is limited to $1s_{\Lambda}$ and 
$1p_{\Lambda}$ states listed in Table~\ref{tab:data}. We fit to such states in 
\textit{one} of the nuclear $1p$-shell hypernuclei listed in the table, where 
the $1s_{\Lambda}$ state is bound by over 10~MeV while the $1p_{\Lambda}$ 
state has just become bound. This helps resolve the density dependence of 
$V_{\Lambda}^{\rm opt}$ by setting a good balance between its two components, 
$V_{\Lambda}^{(2)}(\rho)$ and $V_{\Lambda}^{(3)}(\rho)$, following it all the 
way to \la{208}{Pb} the heaviest hypernucleus marked in Fig.~\ref{fig:mdg}. 
We chose to fit the \lam{16}{N} precise $B^{\rm exp}_{\Lambda}(1s,1p)$ 
values derived, respectively, from the first and third peaks to the left in 
Fig.~\ref{fig:L16N}. The extremely simple $1p$ proton hole structure of the 
$^{15}$N nuclear core in this case removes most of the uncertainty arising 
from spin-dependent residual $\Lambda N$ interactions \cite{Mill08}. The 
fitted optical-potential parameters $b_0$, Eq.~(\ref{eq:V2}), and $B_0$, 
Eq.~(\ref{eq:V3}), are then used to calculate the $B^{1s,1p}_{\Lambda}$ 
values of the other eight species listed in Table~\ref{tab:data}. 

\section{Results} 
\label{sec:res} 

The two strength parameters $b_0,B_0$ of the optical potential terms 
Eqs.~(\ref{eq:V2},\ref{eq:V3}) were obtained by fitting to the \lam{16}{N} 
$B^{\rm exp}_{\Lambda}(1s,1p)$ values listed in Table~\ref{tab:data}. 
Suppressing Pauli correlations by setting $\alpha_P=0$ in 
Eqs.~(\ref{eq:Cpauli1},\ref{eq:Cpauli2}), the resulting $\Lambda$ 
potential depth $D_{\Lambda}=-27.4$~MeV reflects a sizable cancellation 
between a strongly attractive two-body potential depth $D_{\Lambda}^{(2)}$ 
and a strongly repulsive three-body potential depth $D_{\Lambda}^{(3)}$. 
The overall agreement between calculations and experiment is acceptable, 
but some underbinding appears to develop for increasing mass numbers $A$, 
noticed clearly in the three heaviest $1s_{\Lambda}$ and two heaviest 
$1p_{\Lambda}$ states. The resulting $b_0$ is about half of the known 
$\Lambda p$ scattering length of $(1.7\pm 0.1)$~fm \cite{Alex68,HIRES10}. 

\begin{figure}[h]
\centering
\includegraphics[width=0.6\textwidth]{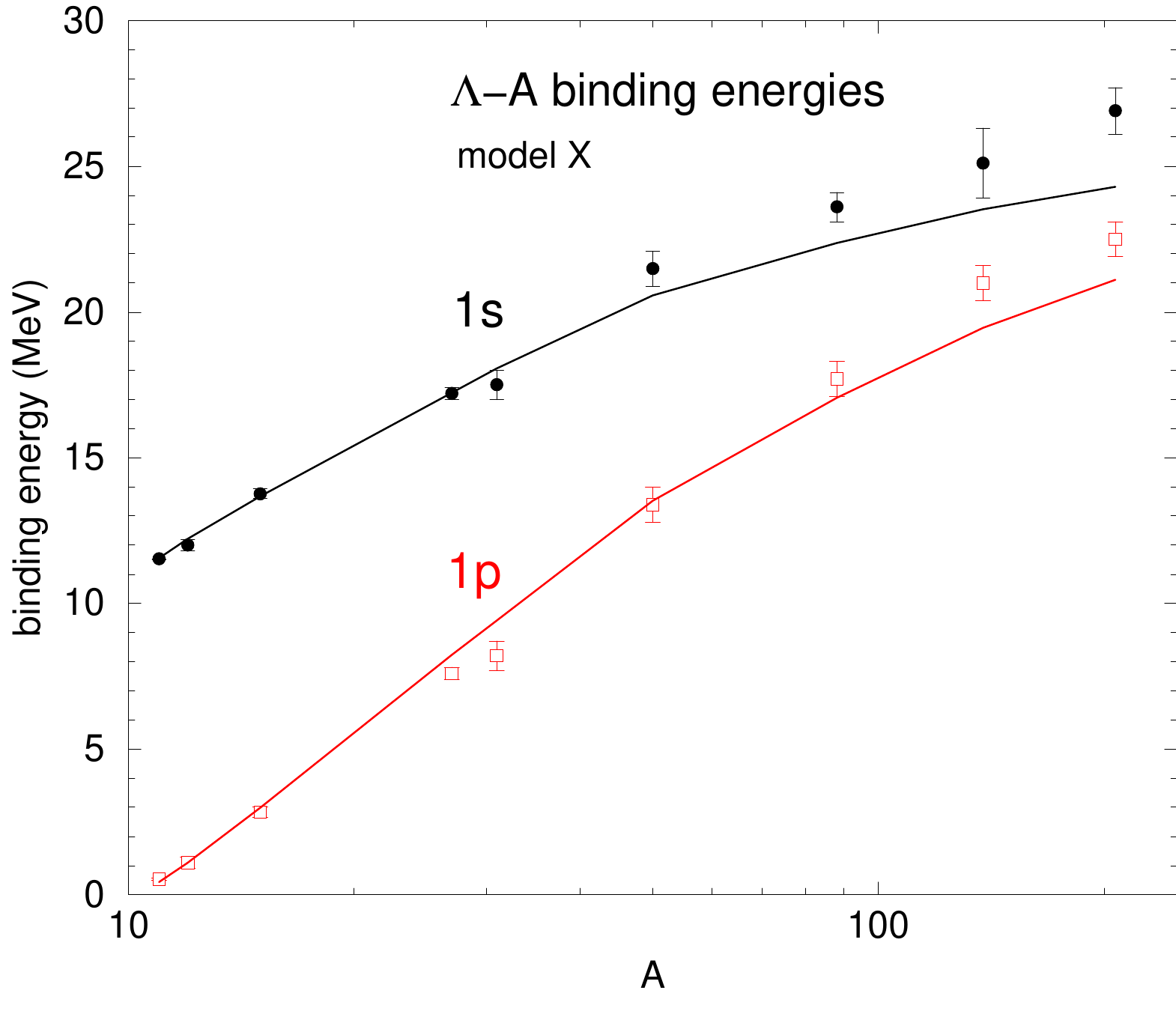} 
\caption{$B_{\Lambda}^{1s,1p}(A)$ values from model X compared with data. 
Continuous lines connect calculated values.} 
\label{fig:modelX} 
\end{figure}

When the full potential Eqs.~(\ref{eq:V2}-\ref{eq:Cpauli2}) is used (marked 
here as model X, including Pauli correlations through $\alpha_P=1$) the 
overall picture remains unchanged regarding underbinding for the heavier 
elements, see Fig.~\ref{fig:modelX}. However, the fit parameter $b_0$=1.85~fm 
agrees now with the $\Lambda p$ scattering length. The other parameter, 
$B_0 = 0.170 $~fm, is about twice smaller than for $\alpha_P=0$. 

\begin{figure}[!t]
\centering 
\includegraphics[width=0.7\textwidth]{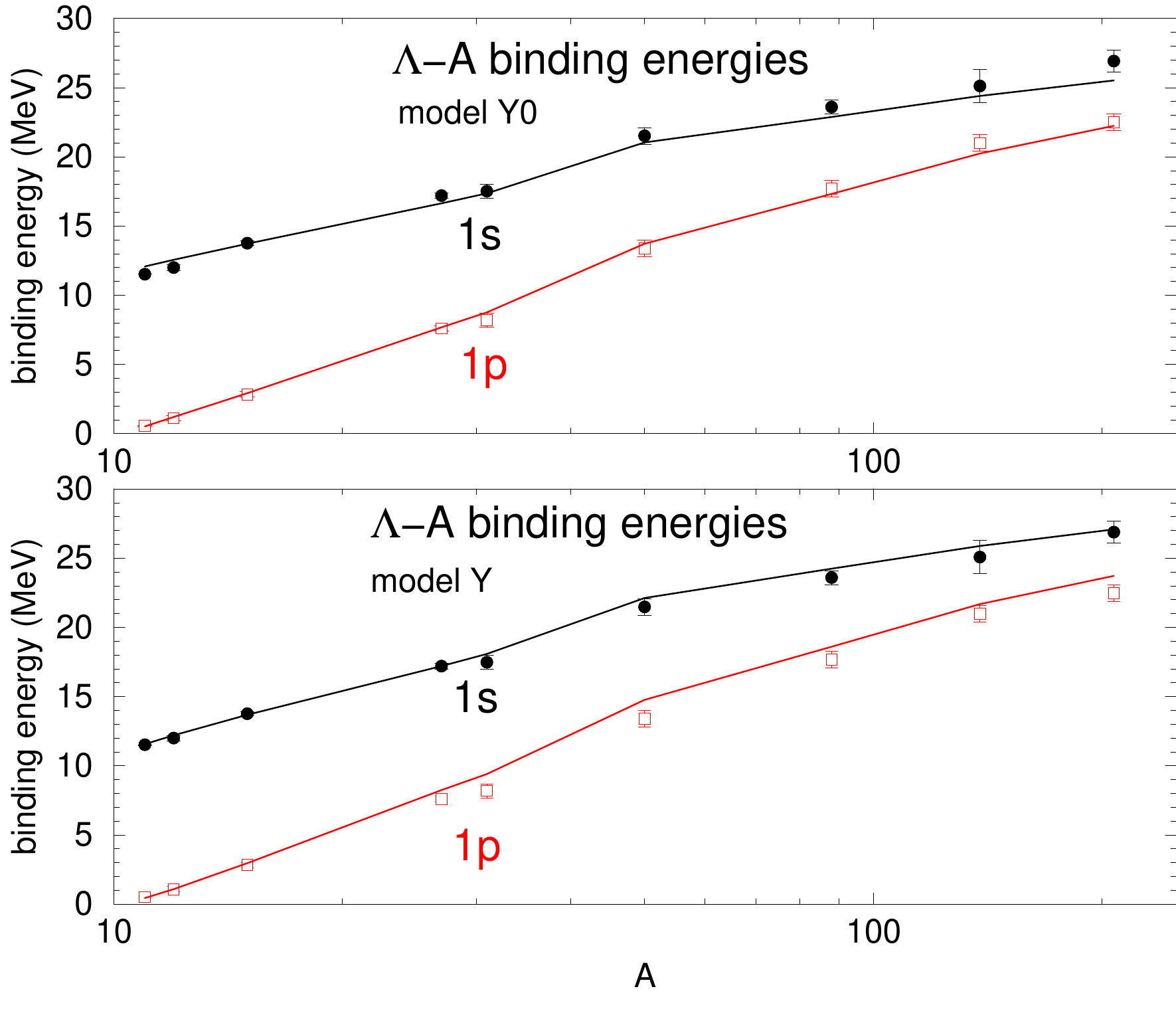} 
\caption{$B_{\Lambda}^{1s,1p}(A)$ values from models Y0 and Y compared with 
data, see text. Continuous lines connect calculated values.} 
\label{fig:modelsY0Y2} 
\end{figure} 

The phenomenon of underbinding associated with the optical potential 
Eqs.~(\ref{eq:V2}-\ref{eq:Cpauli2}) is likely to be a result of the use of 
$\rho ^2$  in nuclei where excess neutrons occupy shell-model orbits higher 
than those occupied by protons. This situation occurs in Fig.~\ref{fig:modelX} 
for the four hypernuclei with $A\gtrsim 50$. Expecting that direct three-body 
$\Lambda NN$ contributions involving one `core' nucleon and one `excess' 
nucleon vanish upon summing on the $T$=0 `core' closed-shell nucleons, 
we modify $\rho^2=(\rho_{\rm core}+\rho_{\rm excess})^2$ by discarding the 
bilinear term $\rho_{\rm core}\,\rho_{\rm excess}$, thereby replacing $\rho^2$ 
in $V^{(3)}_{\Lambda}$, Eq.~(\ref{eq:V3}), by 
\begin{equation}
\rho_{\rm core}^2+\rho_{\rm excess}^2~~=~~(2\rho_p)^2+(\rho_n-\rho_p)^2
\label{eq:exc2}
\end{equation}
in terms of the input densities $\rho_p$ and $\rho_n$. This ansatz is 
consistent with an overall isospin factor $\tau_1\cdot\tau_2$ in two-pion 
exchange $\Lambda NN$ forces, as first realized back in 1958~\cite{Spitzer58}.
Results of applying this ansatz are shown in the lower part of 
Fig.~\ref{fig:modelsY0Y2} as model~Y, where the underbinding of calculated 
1$s_\Lambda$ and 1$p_\Lambda$ binding energies noticed in model X is no longer 
observed. The fit parameters, nevertheless, are the same as for model X above. 
In the upper part of Fig.~\ref{fig:modelsY0Y2}, model Y0 shows similar results 
where the Pauli-correlations correction in model Y, Eq.~(\ref{eq:Cpauli2}), 
is replaced by Eq.~(\ref{eq:Cpauli1}). This provides a rough estimate of the 
impact of $1/A$ corrections typical for our optical-potential methodology. 
Potential depth values in model Y are $D^{(2)}_{\Lambda}=-41.6$~MeV, 
$D^{(3)}_{\Lambda}=13.9$~MeV. 

To estimate uncertainties, we act as follows: (i) decreasing the input value 
of $B_{\Lambda}^{1s}$(\lam{16}{N}) fitted to by 0.2~MeV, thereby getting 
halfway to the central value of $B_{\Lambda}^{1s}$(\lam{16}{O})=(13.4$\pm
$0.4)~MeV for \lam{16}{O}~\cite{Finuda11} the charge-symmetric partner of 
\lam{16}{N}, results in approximately 10\% larger value of $D^{(3)}_{\Lambda}
$, and (ii) applying Pauli correlations to $V^{(3)}_{\Lambda}$ too reduces 
$D^{(3)}_{\Lambda}$ roughly by 10\%. In both cases $D^{(2)}_{\Lambda}$ 
increases moderately by $\lesssim$1~MeV. On the other hand, $D^{(2)}_{\Lambda}
$ decreases by 1.7~MeV if Eq.~(\ref{eq:Cpauli1}) is used for $C_{\rm Pauli}
(\rho)$ instead of Eq.~(\ref{eq:Cpauli2}). Considering these uncertainties, 
our final values are (in MeV)
\begin{equation}
D_\Lambda^{(2)}=-40.6\pm 1.0\,\,\,\,\,\, D_\Lambda^{(3)}=13.9\pm 1.4
\,\,\,\,\,\, D_{\Lambda}=-26.7\pm 1.7 
\label{eq:D} 
\end{equation}

\section{Discussion}
\label{sec:Disc}

The $D^{(2)}_{\Lambda}$ and $D^{(3)}_{\Lambda}$ values in 
Eq.~(\ref{eq:D}) are considerably smaller than those deduced in QMC 
calculations~\cite{LGP13,LPG14}. Note that the QMC nuclear densities 
$\rho_{\rm QMC}(r)$ are much too compact with respect to our realistic 
densities, with nuclear r.m.s. radii $r_N$(QMC) about 0.8 of the known 
r.m.s. charge radii in $^{16}$O and $^{40}$Ca~\cite{Lon13}. Since $\rho$ 
scales as $r_N^{-3}$, applying it to the density dependence of our 
$V^{\rm opt}_{\Lambda}$ would transform $D^{(2)}_{\Lambda}$ and 
$D^{(3)}_{\Lambda}$ of Eq.~(\ref{eq:D}) to as large depth values as 
$D^{(2)}_{\Lambda}$(QMC)=$(-$79.3$\pm$2.0)~MeV and $D^{(3)}_{\Lambda}
$(QMC)=(53.0$\pm$5.3)~MeV, their sum $D_{\Lambda}$(QMC)=$(-$26.3$\pm$5.7)~MeV 
agreeing within uncertainties with ours. 

\begin{table}[htb]
\begin{center}
\caption{$\Lambda$-nuclear potential depths (in MeV) from two SHF calculations 
fitting $B_{\Lambda}$ data points, and from our own $V_{\Lambda}^{\rm opt}
(\alpha_P=0)$ two-parameter ($b_0, B_0$) fit to the two $B_{\Lambda}^{1s,1p}
$(\lam{16}{N}) values listed in Table~\ref{tab:data}.}
\begin{tabular}{ccccc}
\hline
Method & Data Points & $D_{\Lambda}^{(2)}$ & 
$D_{\Lambda}^{(3)}$ & $D_{\Lambda}$ \\
\hline
SHF~\cite{MDG88} & 3 & $-$57.8 & 31.4 & $-$26.4  \\
SHF~\cite{SH14} & 35 & $-$55.4 & 20.4 & $-$35.0  \\
$V_{\Lambda}^{\rm opt}(\alpha_P=0)$~\cite{FG22} & 2 & $-$57.6 & 30.2 & 
$-$27.4 \\
\hline 
\end{tabular}
\label{tab:SHF}
\end{center}
\end{table}

Smaller-size but still inflated values of $D^{(2)}_{\Lambda}$ and 
$D^{(3)}_{\Lambda}$ are obtained by applying the Skyrme Hartree Fock 
(SHF) methodology~\cite{MDG88,SH14}. Apart from small nonlocal potential 
terms and effective mass corrections, the SHF $\Lambda$-nuclear 
mean-field potential $V_{\Lambda}(\rho)$ consists of two terms: 
$V_{\Lambda}^{(2)}(\rho)\propto\rho$ and 
$V_{\Lambda}^{(3)}(\rho)\propto\rho^2$. A large-scale SHF fit~\cite{SH14} 
of the corresponding $\Lambda$ potential depths to 35 $B_{\Lambda}$ data 
points is listed in the middle row of Table~\ref{tab:SHF}. We note that 
the overall $D_{\Lambda}=-35$~MeV value becomes $-$31~MeV upon including 
a $\Lambda$ effective-mass correction, a bit closer to the other 
$D_{\Lambda}$ values listed in the table. Similar results, particularly 
for $D_{\Lambda}^{(2)}$, can be obtained in fact by choosing a considerably 
smaller number of fitted data points, as shown by the fits listed in the other 
two rows of the table. The 11~MeV difference between the $D_{\Lambda}^{(3)}$ 
values derived in these two SHF calculations arises mostly from nonlocal 
lower-power density terms, like $\rho^{5/3}$, present in~\cite{SH14} but 
absent in~\cite{MDG88}. Interestingly, the last row lists a fit to the two 
$B_{\Lambda}^{1s,1p}$(\lam{16}{N}) values used here when Pauli correlations 
are suppressed, $\alpha_P=0$ in Eq.~(\ref{eq:Cpauli1}). The sizable difference 
between $D_{\Lambda}^{(2)}$ and $D_{\Lambda}^{(3)}$ values listed in 
Table~\ref{tab:SHF}, all of which disregard Pauli correlations, and the 
$V_{\Lambda}^{\rm opt}$ values listed in Eq.~(\ref{eq:D}) demonstrates the 
importance of including in $V_{\Lambda}^{\rm opt}$ a Pauli-correlations term 
($\alpha_P=1$) starting as $\rho^{4/3}$.

\section{Summary} 
\label{sec:Sum} 

In summary, we have presented a straightforward optical-potential analysis 
of $1s_\Lambda$ and $1p_\Lambda$ binding energies across the periodic table, 
$12\leq A\leq 208$, based on nuclear densities constrained by charge r.m.s. 
radii. The potential is parameterized by constants $b_0$ and $B_0$ in front 
of two-body $\Lambda N$ and three-body $\Lambda NN$ interaction terms. These 
parameters were fitted to precise $B^{\rm exp}_{\Lambda}(1s,1p)$ values in 
\lam{16}{N}~\cite{JLab09} and then used to evaluate $B^{1s,1p}_{\Lambda}$ 
values in the other hypernuclei considered here. Pauli correlations were 
found essential to establish a correct balance between $b_0$ and $B_0$, 
as judged by $b_0$ coming out in the final model Y analysis close to 
the value of the $\Lambda N$ spin-averaged $s$-wave scattering length. 
Good agreement was reached in this model between the calculated 
$B^{1s,1p}_{\Lambda}$ values and their corresponding $B^{\rm exp}_{\Lambda}$ 
values, see Fig.~\ref{fig:modelsY0Y2}. 

The potential depth $D^{(3)}_{\Lambda}$ derived here, Eq.~(\ref{eq:D}), 
suggests that in symmetric nuclear matter the $\Lambda$-nucleus 
potential becomes repulsive near three times $\rho_0$. Our derived 
depth $D^{(3)}_{\Lambda}$ is larger by a few MeV than the one yielding 
$\mu(\Lambda) > \mu(n)$ for $\Lambda$ and neutron chemical potentials 
in purely neutron matter, respectively, under a `decuplet dominance' 
construction for the underlying $\Lambda NN$ interaction terms within 
a $\chi$EFT(NLO) model \cite{GKW20}. This suggests that the strength of 
the corresponding repulsive $V_\Lambda^{(3)}$ optical potential component, 
as constrained in the present work by data, is sufficient to prevent 
$\Lambda$ hyperons from playing active role in neutron-star matter, 
thereby enabling a stiff EoS that supports two solar-mass neutron stars.

\section*{Acknowledgments}
\begin{acknowledgement}
One of us (A.G.) thanks Ji\v{r}\'{i} Mare\v{s} and other members of the 
HYP2022 organizing team for their generous hospitality during the Conference. 
The present work is part of a project funded by the European Union's Horizon 
2020 research \& innovation programme, grant agreement 824093.
\end{acknowledgement}

\end{document}